\documentclass[10pt]{article}
\usepackage{authblk}
\usepackage{alltt,graphicx,amssymb,lineno}
\usepackage{amsmath,mathrsfs,graphicx,amssymb}
\usepackage{color}
\usepackage{natbib}
\newcommand{\ba}{\begin{eqnarray}}
\newcommand{\ea}{\end{eqnarray}}
\renewcommand{\rho}{\varrho}
\newcommand{\eps}{\varepsilon}

\def\bS{{\bar S}}

\def\ev{{\bf e}}
\def\s0{{S_0}}

\def\nv{{\bf n}}

\def\tv{{\bf t}}
\def\fv{{\bf f}}
\def\rv{{\bf r}}

\def\tv{{\bf t}}

\def\F{{\rm F}}
\def\E{{\rm E}}

\def\Fv{{\bf F}}

\def\Mv{{\bf M}}

\def\Tv{{\bf T}}

\def\bs{\bar{S}}
\def\bt{\bar{\theta}}

\def\t0{{\theta_0}}

\def\de{{\rm d}}

\def\eps{\epsilon}

\begin{document}

\title{Snap buckling of a confined thin elastic sheet}
\author[1]{\normalsize G. Napoli}
\author[2]{S. S. Turzi}
\affil[1]{\small Dipartimento di Ingegneria dell'Innovazione, Universit\`a del Salento, via per Monteroni, Edificio 'Corpo O', 73100 Lecce, Italy }

\affil[2]{\small Dipartimento di Matematica, Politecnico di Milano, Piazza Leonardo da Vinci 32, 20132 Milano, Italy}

\date{ }

\maketitle

\abstract{
A growing or compressed thin elastic sheet adhered to a rigid substrate can exhibit a buckling instability, forming an inward hump. Our study shows that the strip morphology depends on the delicate balance between the compression energy and the bending energy. We find that this instability is a first order phase transition between the adhered solution and the buckled solution whose main control parameter is related to the sheet stretchability. In the nearly-unstretchable regime we provide an analytic expression for the critical threshold. Compressibility is the key assumption which allows us to resolve the apparent paradox of an unbounded pressure exerted on the external wall by a confined flexible loop.
}

\section{Introduction}
Elastic strips and rods are fundamental physical structures that can be found in many contexts on many different length-scales.  Often, they appear in the theoretical description of physical phenomena where the interplay between physics and geometry plays a key role.  The collapse of an elastic sheet adhered to a curved substrate is certainly one of them.  This discontinuous buckling instability appears in various problems in biology (shapes and morphogenesis of the mitochondrion \citep{Sakashita:2014}), physics (packing of thin films \citep{Boue:2006},  delamination of thin elastic films \citep{Vella:2009, Wagner:2013}, design of stretchable electronic devices \citep{Sun:2006}, elastocapillary snapping \citep{Fargette:2014})  and engineering (collapse of buried steel pipelines \citep{Omara:1997,Vasilikis:2009}).  

A simple example illustrates the physical phenomenon we are concerned with. Let us imagine to insert a rectangular piece of paper inside a rigid cylindrical substrate. We hold the two edges with the hands so that the sheet of paper is kept in contact with the cylinder. Subsequently, we apply an increasing tangential compressive force. Due to presence of the surrounding container, the sheet of paper cannot freely deform to the outside. Initially it appears to be undeformed or slightly compressed. However, when the applied force crosses a suitable threshold the strip abruptly buckles and forms an inward hump. What is the critical force at which the collapse occurs? How are the geometric features of the hump at the transition related to the material constants and to the geometry of the container? Although the buckling of slender elastic structures is one of the oldest fundamental problems, nobody seems to have addressed these points. Furthermore, it is worth noticing that classical bifurcation analysis 
fails in this problem due to the presence of the confinement.

A strictly related problem deals with the packing of a flexible cylinder inside a rigid circular tube of smaller radius. This is widely explored in the literature, both theoretically \citep{Cerda:2005} and experimentally \citep{Boue:2006}. Recent papers have also included the possibility of adherence by capillarity \citep{DePascalis:2014, Rim:2014} or the extension to the spherical geometry \citep{Kahraman:2012}.  In the usual theoretical treatment, the strip is modelled as an inextensible Euler beam, whose energy is only associated to the bending mode. The inextensibility assumption, first suggested by \cite{Rayleigh}, is based on energetic considerations. The bending energy and the stretching energy scale differently with the beam thickness. When the structure thickness is diminished without limit, stretching is energetically prohibitively expensive and the distortion is pure bending.  However, the validity of this assumption is challenged when the thickness becomes comparable to some other length-
scales in the problem. {Moreover, axial compressibility is known to change the nature of the phase transition. For example, the usual supercritical bifurcation point in the classical inextensible Euler beam becomes a subcritical bifurcation \citep{01magn,Audoly:2010}.

However in our confined problem, the assumption of an axially inextensible beam shows its inadequacy in a more fundamental way. For instance, in the limit of arbitrarily small hump, the inextensible Euler's Elastica model predicts the unphysical result of an infinite pressure exerted by the beam on the external container \citep{Cerda:2005, DePascalis:2014}. This prevents the transition to a buckled state as it would require an infinite force to induce the buckling.}

{To avoid this unacceptable behaviour, in this paper we revise this effect by allowing the sheet to both stretch and bend.  Within this framework, the energy landscape exhibits a first order phase transition between a state where compression energy dominates, to another where  bending energy prevails.  From the mechanical point of view, a {\it snap through} buckling occurs: as the sheet grows, it suddenly switches from a completely adhered solution to a buckled solution showing a symmetric inward hump. The transition is studied numerically and, in the limit of {\it weak stretchability}, we provide the analytic expressions for the main quantities at the critical point. Interestingly, we find that for homogeneous beams the threshold does not depend on the material parameters but only on the geometric features of the problem.}

\section{The model}
We assume a planar configuration so that the longitudinal profile of the strip can be modelled as a stretchable and flexible rod. This is represented by a parametric curve $\rv(S)$, with $S\in [-L/2,L/2]$, where $L$ denotes the length of the strip in the stress-free configuration and $S$ is the referential arclength. The governing nonlinear Kirchhoff equations express the balance of linear and angular momentum. When an external distributed force $\fv$ is applied and in absence of external distributed torques, they read
\ba
\Tv' + \fv ={\bf 0},
\label{b1}
\ea
\ba
\Mv' + \rv' \times \Tv = {\bf 0},
\label{b2}
\ea
where $\Tv(S)$ and $\Mv(S)$ represent the internal forces and the internal torques, respectively. The prime denotes differentiation with respect to $S$; hence, $\rv' = \lambda \tv$, where $\lambda$ is the local stretch and $\tv$ is the unit tangent. The inextensibility condition corresponds to $ \lambda = 1$.  

Balance equations are complemented by the linear constitutive laws \citep{Dill:1992}:
\ba
\Tv = b (\lambda -1) \tv + T_n \nv,
\ea
\ba
\Mv = k \tv \times \tv' ,
\label{momento}
\ea
where the positive constants $b$ and $k$ represent the stretching and the bending rigidity, respectively.  The quantity $\ell = \sqrt{k/b}$ defines an intrinsic characteristic length. Since $b= E A$ and $k = E I$, where $E$ is the elastic modulus of the material, $A$ is the area of the film cross-section and $I$ is the second moment of area of the strip cross-section, the intrinsic length $\ell$ turns out to be of the order of the strip thickness. Only the tension is given constitutively; the shear internal force  $T_n$ is related to the derivative of the internal torque by Eq. \eqref{b2}. We  parametrize the tangent and the normal unit vectors by
\[
\tv = \cos \theta(S) \ev_x + \sin \theta(S) \ev_y, \qquad \nv = -\sin \theta(S) \ev_x + \cos \theta(S) \ev_y
\]
and, hence, 
$
\ev_z = \tv \times \nv.
$
Consequently, Eqn. \eqref{momento} reduces to
$
\Mv = k \theta' \ev_z.
\label{c2}
$

\begin{figure}[ht]
\centerline{\includegraphics[scale=1.2]{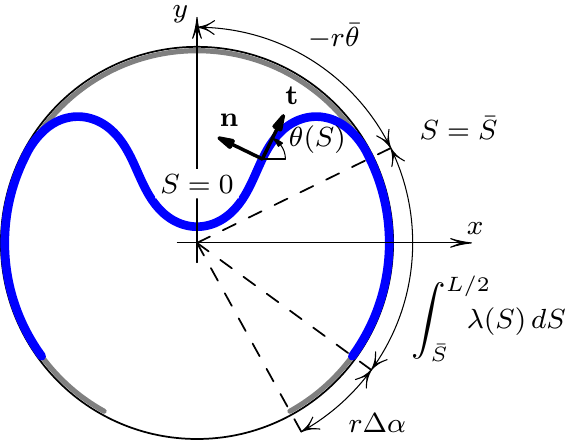} }
\caption{\label{blister} Schematic representation of the strip deformation. The gray curve represents the referential configuration, while the blue curve shows the buckled configuration with a hump due to the displacement $r\Delta \alpha$ of the end points.}
\end{figure}

Furthermore, we posit a symmetric equilibrium configuration of the rod and assume that the \emph{displacement} of its end-points can be controlled: the point with referential coordinate $S=L/2$ undergoes a tangential compressive displacement $r \Delta \alpha$,  $\Delta \alpha \geq 0$. The strip can both stretch and bend, therefore there are at least two equilibrium configurations with the same imposed $\Delta \alpha$:  (i) the {\it adhered solution} where the strip, uniformly compressed, totally adheres to the wall, and (ii)  the {\it collapsed solution} with an inward symmetric hump. The adhered solution is a uniformly compressed arc of circumference with
\[
\lambda_{adh}= 1 - 2  r \Delta \alpha/L, \qquad \theta'_{adh} = - \lambda_{adh}/r.
\]
On the other hand, the collapsed solution comprises two parts: a free (non-adhered) curve for $S \in (-\bS, \bS)$ and two adhered pieces for $S \in[-L/2,-\bS]  \cup [\bS, L/2]$ (see Figure \ref{blister}). The symmetry of the strip implies that the function $\theta(S)$ is odd and allows us to restrict the study of the solution in the range $S \in [0,L/2]$.

Since the adhered part, $S \in [\bS,L/2]$, is an arc of circumference with $\theta' (S)=-\lambda/r$, the internal torque is $\Mv = -k (\lambda/r) \ev_z$. Therefore, the equilibrium equations for this imposed geometry necessarily give $T_n=0$ and $\lambda'  = 0$, whence $\lambda(S) = \bar{\lambda}= {\rm const}$.  However, the compressive tangential force $\Fv= -F \tv$ for a given $\Delta \alpha$ is not known \emph{a-priori} but it is related to the stretch by $F = b(1-\bar \lambda)$.

In the non-adhered region, $S \in [0,\bS]$, there is no applied distributed load ($\fv= {\bf 0}$). Furthermore, since $\theta(S)$ is odd, the vertical component of the tension vanishes, while the horizontal component, denoted by $h$, is constant and satisfies the  identity
$
b(\lambda-1) = h \cos \theta.
$
The balance of the torque yields the second order differential equation for the free part
\ba
\theta'' +  \lambda \tau \sin \theta =0.
\label{eul}
\ea
where $\tau = - h/k$. 

The deflection angle $\theta(S)$ should satisfy the Dirichlet  boundary conditions $\theta(0)=0, \; \theta(\bS) = \bt$. Further boundary  conditions  are provided by requiring the continuity of the tangential component of the internal force and the continuity of the torque at the detachment point  $S=\bS$. These yield 
\[
h \cos \bt =-b(1-\bar \lambda), \qquad \theta'(\bS) = - \bar \lambda r^{-1},
\] respectively. One more constraint translates the geometrical condition that the detachment point must lie on the circumference of radius $r$
\ba
\int_0^ {\bS} \lambda \cos \theta \de S = - r \sin \bt.
\label{vinco}
\ea
This establishes a relationships between $\bS$ and $\bt$. The displacement of the end-point, $r \Delta \alpha$, is related to $\bar\lambda$, $\bs$ and $\bt$ by (Figure \ref{blister})
\ba
r\Delta \alpha  = r \bt + (1-\bar \lambda) \frac{L}{2} + \bar \lambda \bs. 
\label{geo1}
\ea
{This geometrical identity is simply obtained by requiring that the total length is the sum of the hump length and the adhered length.}
\section{Results}
{We start by observing that} for a perfectly unstretchable strip the compression modulus diverges and the intrinsic length vanishes, $\ell=0$. The tangential component of the internal force, $T_t$, then becomes a Lagrange multiplier associated to the inextensibility constraint $\lambda=1$ and the adhered solution is admissible only for $\Delta \alpha=0$.

By contrast, when both compression and bending are allowed, the energy must attain its absolute minimum in the observed solution, either  adhered or collapsed, which is then expected to be stable. The other possible solution is either non-existing, unstable or metastable, depending on the energy landscape. The total elastic energy consists of two terms, related respectively to the bending mode and to the stretching (or compression) mode: 
\ba
W = \frac{k}{2}\int_{-L/2}^{L/2} (\theta')^2 \de S  + \frac{b}{2}\int_{-L/2}^{L/2} (1-\lambda)^2 \de S. 
\label{energia}
\ea
Equation \eqref{energia} assumes a particularly simple analytic form when evaluated in the adhered solution
\ba
W_{adh} = \frac{1}{2} \frac{k L}{r^2} + 2 \frac{k}{r} \Delta \alpha \left[- 1 +   \frac{r}{L} (1+\xi^{-2}) \Delta \alpha \right],
\label{en_adh}
\ea
where $\xi = \ell/r>0$  measures the compressibility strength at fixed radius. 
\begin{figure}[ht]
\centerline{\includegraphics[scale=1]{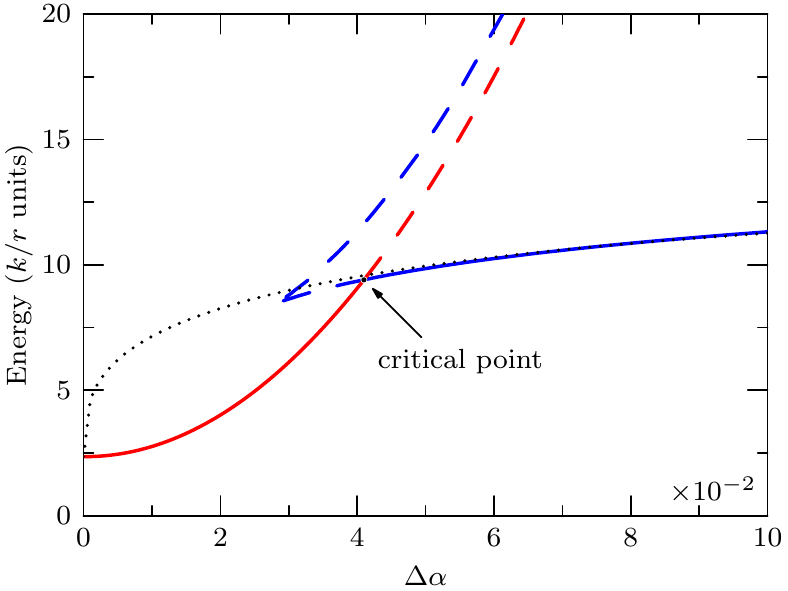}}
\caption{\label{figura_energie} {Plot of the equilibrium free energy $W$ for the adhered solution (red) and the collapsed solution (blue) with $L=\tfrac{3}{2}\pi r$ and $\xi=0.01$. Solid lines represent solutions with least energy; dashed lines represent metastable states or maxima. For small $\Delta \alpha$ only the adhered solution is admissible. The dotted line represents the energy of a perfectly inextensible strip as given in Eq.\eqref{eq:W_approx}.}}
\end{figure}

On the other hand, the energy associated with the collapsed solution, $W_{col}$, is not amenable for a simple analytical approximation and is best calculated with a numerical simulation. Figure \ref{figura_energie}  shows the comparison between the energies associated with each branch as a function of $\Delta \alpha$, for fixed $\xi=0.01$. The adhered branch depends quadratically on $\Delta \alpha$ (see equation \eqref{en_adh}). The collapsed solution comprises two branches. The upper branch, with higher energy,  corresponds to small humps, while the lower branch corresponds to equilibrium solutions with larger humps. Finally, it is worth noticing that for sufficiently small values of $\Delta \alpha$ only the adhered solution is admissible.

\begin{figure}[ht]
\centerline{\includegraphics[scale=1]{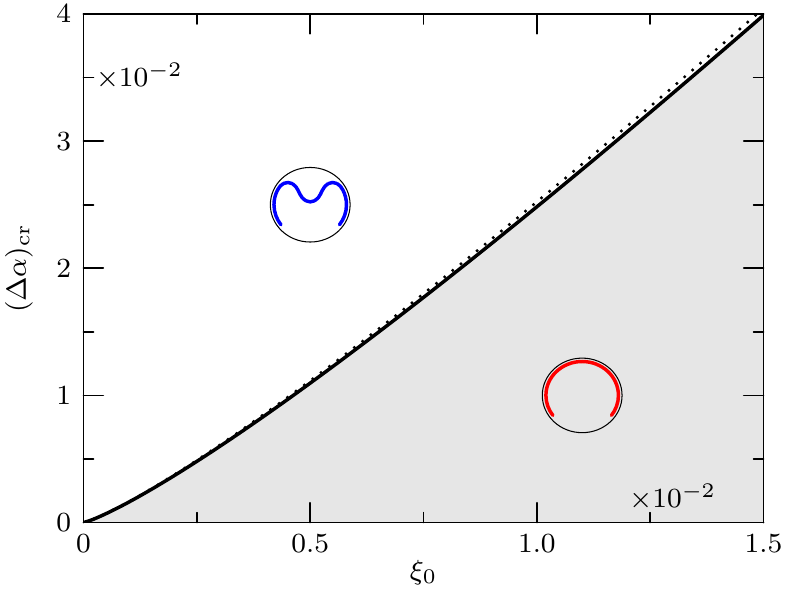}}
\caption{\label{figura_dacritico} {Critical threshold $(\Delta \alpha)_{cr}$ as a function of the reduced compressibility. The solid line represents the numerical results, while the dotted line represents the analytical approximate result as given in Eq.\eqref{soglia_critica}. Typical shapes are shown in the inset.}}
\end{figure}

Figure \ref{figura_energie} clearly shows the first-order transition that occurs between the two solutions. For small displacements of the end-points, the adhered solution is energetically favoured. However, when the displacement crosses a critical value $r(\Delta \alpha)_{cr}$, the energy attains its absolute minimum at the collapsed solution \emph{with larger hump}.  Therefore, as $\Delta \alpha$ is increased, the strip undergoes a discontinuous transition, abruptly passing from an adhered configuration to a buckled solution with an inward hump. The presence of compressibility  implies a lower bound for the hump size: humps whose dimensions are small compared to the characteristic length $\ell$ cannot be observed and the strip is simply compressed. The critical threshold increases as the strip becomes softer (see Figure \ref{figura_dacritico}). In particular, in the inextensible limit the critical threshold vanishes, as expected.

In the {\it nearly unstretchable} regime $(\xi \ll 1)$ one can estimate the critical threshold analytically. Indeed, at low $\xi$ and close to the critical point, it has been observed numerically that the collapsed solution is nearly equivalent to the buckled profile of a perfectly inextensible strip. Figure \ref{figura_energie} clearly shows this agreement in terms of elastic energies. From the analytic approximation we also learn that the relevant variable for scaling the critical threshold is the {\it reduced compressibility}, defined as 
\[
\xi_0 := \xi \sqrt{L/2r}.
\]
This indicates an unexpected non-trivial dependence on the geometrical and material parameters. {For homogeneous materials $\xi_0$ is just a geometrical dimensionless parameter, since both the compression and the bending rigidity are linear functions of the Young modulus. By contrast, the material moduli could play a role in the case of composite materials where the dependence of $b$ and $k$ on the Young modulus is more complex.}

When the compression modulus is large ($\xi \ll 1$), compression becomes energetically prohibitively expensive for small $\Delta\alpha$. The strip then prefers to buckle and pay some bending energy provided it can relax its compression energy. Therefore, it is natural to approximate the collapsed strip as inextensible. The advantage of this approximation is that it is now possible to perform an analytic treatment. In particular,  Eq.\eqref{eul}  simply reduces to the nonlinear pendulum equation with first integral
\begin{equation}
(\theta')^2 = 2\tau(\cos \theta - \cos \theta_0) , 
\label{eq:first_integral}
\end{equation}
where $\theta_0 = \theta(\s0) \in [0,\pi]$ is the maximum value of $\theta(S)$ in $(0,\bs)$. 
The boundary condition for $\theta'$ at $\bs$ becomes $\theta'(\bs)=-1/r$. This is used together with the first integral to eliminate $\tau$ in favour of $\theta_0$. Thus, we arrive at the first order differential equation 
\begin{equation}
\theta' =  \pm \frac{1}{r} \sqrt{\frac{\cos \theta - \cos \theta_0}{\cos \bt - \cos \theta_0}},
\label{tetas}
\end{equation}
where the sign $+$ (respectively, $-$) is to be used in the interval $S\in(0,\s0)$ (respectively, $S\in(\s0,\bs)$). By symmetry $\theta(0)=0$ and Eq.\eqref{tetas} evaluated at $S=0$ shows that $\cos\bt -\cos\t0 >0$. This gives a restriction on the possible values of $\bt$: $|\bt|<\t0$. Furthermore, Eq.\eqref{tetas} is an ordinary differential equation which can be solved by separation of variables in $(0,\bs)$ (see \cite[Sect. 3]{DePascalis:2014})
\begin{equation}
2\F(q_0)-\F(\bar q) = \frac{\bs}{2r} {\sqrt \frac{1-\cos\theta_0}{\cos \bt - \cos \theta_0}},
\label{prima}
\end{equation}
where $\F$ denotes the incomplete elliptic integral of first kind \citep{Abramowitz:1970} and, for ease of notation, we set
\[
q_0 := \{{\t0}/{2}, \csc^2 ({\t0}/{2}) \}, \qquad \bar q := \{{\bt}/{2}, \csc^2 ({\t0}/{2}) \}.
\]
Similarly, we simplify Eq.\eqref{vinco} (with $\lambda=1$) as follows
\begin{eqnarray}
2   \E (q_0)-  \E (\bar q) = - \frac{\bs \cos \t0 + r\sin \bar \theta}{2 r \left(1-\cos\theta_0\right)} \sqrt\frac{{1-\cos\theta_0}}{\cos \bt - \cos \theta_0},
\label{seconda}
\end{eqnarray}
where $\E$ represents the incomplete elliptic integral of second kind. Equations \eqref{prima}, \eqref{seconda}, and the inextensible version of Eq.\eqref{geo1}, $r \Delta \alpha = r \bt + \bs $, provide $\bs$, $\t0$ and $\bt$ as functions of $\Delta \alpha$. In the limit $\Delta \alpha \ll 1$, after some tedious but straightforward calculations which we omit for brevity, it is possible to find the asymptotic expansions of $\t0$ and $\bt$ {\citep{DePascalis:2014}
\begin{align}
\t0 & \approx 2.3454 (\Delta\alpha)^{1/3} - 0.78762 \Delta\alpha + 0.68361 (\Delta\alpha)^{5/3} , \\
\bt & \approx -2.2894 (\Delta\alpha)^{1/3} + 0.83072 \Delta\alpha - 0.62489 (\Delta\alpha)^{5/3}.
\end{align}
The energy of an inextensible buckled solution has a particularly simple expression when it is written in terms of $\t0$, $\bt$ ans $\bs$. To this end, we insert Eq.\eqref{tetas}, Eq.\eqref{vinco} and $\lambda=1$ into Eq.\eqref{energia} to get
\begin{align}
W_{col} & = k \int_{0}^{\bs } (\theta')^2 \de S + k \int_{\bs}^{L/2} (\theta')^2 \de S
= \frac{k}{r^2} \int_{0}^{\bs } \frac{\cos\theta(S) - \cos\t0}{\cos\bt - \cos\t0} \de S + \frac{k}{r^2} \Big(\frac{L}{2}-\bs\Big) \notag \\
& = \frac{1}{2}\frac{k L}{r^2} - \frac{k}{r^2}\frac{\bs \cos\bt + r \sin \bt}{\cos\bt-\cos\t0}.
\label{eq:IncBuckled}
\end{align}
The asymptotic expansions of $\t0$ and $\bt$ are then inserted into Eq.\eqref{eq:IncBuckled} to yield the approximate expression for the energy of the collapsed solution
\ba
W_{col}\! \approx \!\frac{1}{2}\frac{k L}{r^2} \!+ \!\frac{k}{r}\big(w_0 (\Delta \alpha)^{1/3} 
+ w_1 \Delta \alpha + w_2   (\Delta \alpha)^{5/3} \big),
\label{eq:W_approx}
\ea
where $w_0 \approx 23.113, \; w_1 \approx -18.532, \; w_2 \approx 1.6131$. }
Thus, we are able to give an approximate analytic solution of the transition equation $W_{col} =W_{adh}$, which to leading order reads
\begin{equation}
(\Delta \alpha)_{cr}  \approx 6.5813 \; \xi_0^{6/5}.
\label{soglia_critica}
\end{equation}
As shown in Figure \ref{figura_dacritico}, this expression provides a very good account of the numerical results, in the range of compressibility considered.

\begin{figure}[t]
\centerline{\includegraphics[scale=1]{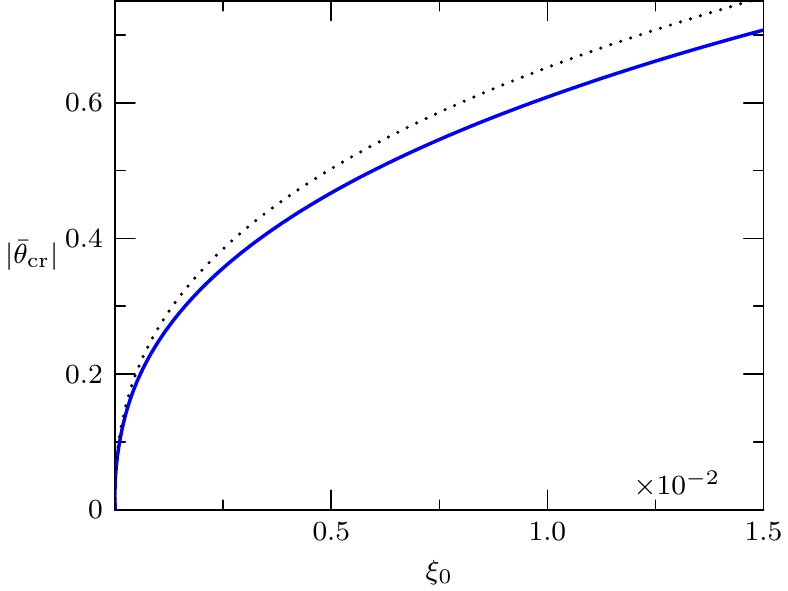}}
\caption{\label{fig_tb_xi} Detachment angle at the transition as a function of $\xi$. The dotted line represents the asymptotic approximation.}
\end{figure}

Figure \ref{fig_tb_xi} sketches the critical behaviour of the detachment angle. This angle is directly related to the hump dimension and is easily accessible experimentally. {For a given $\xi$, $|\bt_{cr}|$ represents the minimum stable detachment angle. Lower values of $\bt$ corresponds to metastable or unstable solution. Thus, we can conclude that there is a limiting size below which the hump cannot develop. The dotted line in Figure \ref{fig_tb_xi} represents the asymptotic approximation}
$ |\bt_{cr}| \approx 4.29 \; \xi_0^{2/5}. $

\begin{figure}[t]
\centerline{\includegraphics[scale=1]{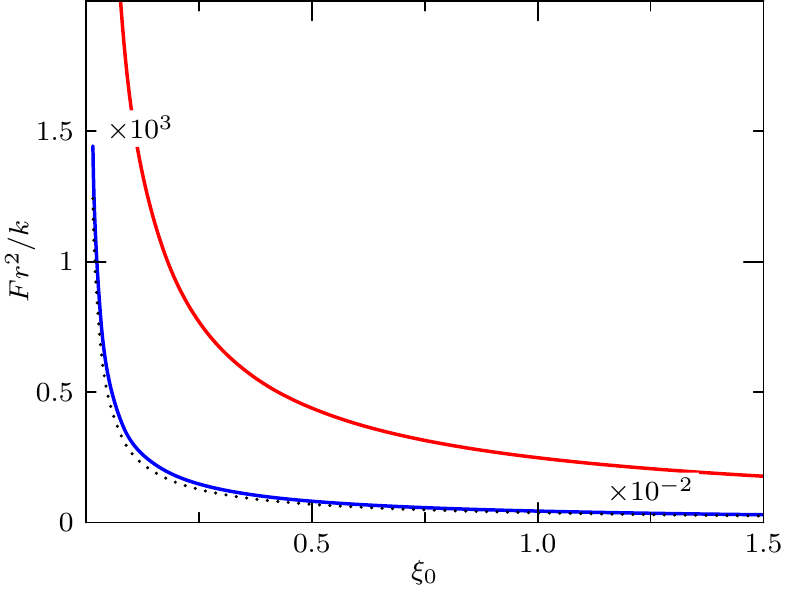}}
\caption{\label{force} Tangential compressive force at the transition in the adhered solution (red line) and collapsed solution (blue line) as a function of $\xi$. The tangential force has a finite jump at the transition. The dotted lines show the asymptotic approximations.}
\end{figure}

The lower bound for the blister size corresponds to a higher bound for the applied tangential force and thus for the pressure exerted on the delimiting container. Figure \ref{force} reports the behaviour at the transition of the tangential force for the adhered (red line) and collapsed (blue line) solutions, as functions of the reduced compressibility. Since at the transition the solution passes from one to the other, the tangential force undergoes a discontinuous jump. This is consistent with the idea of buckling as the mechanism through which the system relaxes its internal stress. {For fixed $\xi_0$, the correspondent value of the red line shows the maximum internal force acceptable by the system under pure compression. The critical load at which the buckling occurs is given by  \mbox{$F_{cr} = 2 b r (\Delta \alpha)_{cr}/L$}. Furthermore, when $\Delta \alpha$ is below its critical threshold the internal force is simply proportional to the strain and hence it vanishes as $\Delta \alpha$ tends to zero}.

\section{Concluding remarks}
In many applied contexts it is useful to study the slightly different problem where the sheet is closed, {\it i.e.} the lateral ends of the strip are glued together. For instance, the elastic sheet can be made of a growing biological membrane confined by a rigid cylinder of radius $r$. In this case, the circumferential growth, suitably measured by $\eps:= (L -  2 \pi r)/(2 \pi r)$, may trigger the instability. For moderate growth, the excess material can simply result in a uniform compression of the adhered strip and in an increase of the hoop stress. By contrast, a further increase of the growth (when the critical threshold has been crossed) leads to the buckling of the membrane and the sheet is then only in partial contact with the container. Our results can be simply adapted to this closed problem. As expected we again find that there is a critical growth $\eps_{cr}$ above which the delamination is preferred with respect to the adhered solution.  Indeed, the equations governing the post-buckling behaviour 
are 
the same except for Eq.\eqref{geo1} that is to be replaced with
\[
\pi r \eps = r \bt + (1-\bar \lambda) {L}/{2} + \bar \lambda \bs.
\]
It is worth noticing explicitly that even the energy expression \eqref{energia} remains unchanged. By contrast, the adhered solution is characterized by a uniform stretch $\lambda_{adh} = 1/(1 + \eps) $ and its elastic energy is 
\[
W_{adh}= \pi k r^{-1} (1 + \eps)^{-1}(1+ \eps^2 \xi^{-2}).
\]
However, it turns out that the two problems are mathematically equivalent provided we substitute $\Delta \alpha$ with $\pi \eps$. We finally find the following approximation for the critical growth: $\eps_{cr} \approx 41.0912\; \xi^{6/5}$. 

As an application of our results, we look at the experiment on the packing of flexible films reported in \cite{Boue:2006}.  {In the first part of their paper, the Authors studied the pressure exerted by the sheet on the external container as a function of $\eps$, here named {\it the confinement parameter}. Their experimental data exhibit an increasing pressure as the confinement parameter $\eps$ is decreased. Numerical studies of the elastica model indicate indeed a divergent pressure in the limit of $\eps \rightarrow 0$. This is also consistent with the theoretical results \citep{Cerda:2005,DePascalis:2014}, which in addition show an asymptotic dependence of the internal forces, and hence the pressure, on $\eps^{-2/3}$}.   However, the thickness of the sheet used in the experiment is $h = 0.1\:$mm, while the container radius is $r=26\;$mm. Since for a rectangular section $\ell = h/\sqrt{12}$, we obtain that $\xi = 1.11 \times 10^{-3}$ and $\eps_{cr}=1.17 \times 10^{-2}$. Unfortunately, 
the smallest experimental value of $\eps$ reported in Figure 3 of \cite{Boue:2006} is just above this critical value and hence they are not able to see the transition experimentally. {On the one hand, this justify the validity of the Elastica model above the critical threshold}. On the other hand, our results predict that, by further decreasing $\eps$, the hump should suddenly disappear. At the transition, the pressure (per unit length) on the container undergoes a jump to $P_{cr} \approx b r^{-1} \eps_{cr} $, while below $\eps_{cr}$ the pressure is proportional to $\eps$, hence it goes to zero as $\eps \to 0$. This solve the apparent paradox, described in \cite{Cerda:2005} , \cite{DePascalis:2014} and \cite{Boue:2006}, of a diverging pressure in the low confinement regime. 

In conclusion, the present analysis recognizes the key role played by stretchability for a correct description of the mechanical collapse of a loaded or growing elastic thin sheet adhered to a curved substrate. Even when the sheet is nearly unstretchable, the energetic contribution of compression cannot be neglected when the length scales in the problem are comparable with the sheet thickness (i.e., small humps). Our model yields analytic approximations for the critical threshold and the minimum size for the humps. {This threshold depends on the reduced compressibility that, for homogeneous materials, is a purely geometrical dimensionless parameter}. Furthermore, it sets an upper bound to the exerted pressure on the container. 

Future works can consider the effects of other physical quantities (i.e. capillary adhesion, intrinsic curvature and gravity) or the coupling of growth with internal stress \citep{Tiero:2014} on the morphology and the critical parameters.
 
 \section*{Acknowledgements} 
 
This research has been carried out within the Young Researchers Project ``Collasso Meccanico di Membrane Biologiche Confinate'', supported by the Italian `Gruppo Nazionale per la Fisica Matematica' (GNFM). 

\end{document}